\documentclass[pdflatex,referee,sn-basic]{sn-jnl}

\usepackage{graphicx}%
\usepackage{multirow}%
\usepackage{amsmath,amssymb,amsfonts}%
\usepackage{amsthm}%
\usepackage{mathrsfs}%
\usepackage[title]{appendix}%
\usepackage{xcolor}%
\usepackage{textcomp}%
\usepackage{manyfoot}%
\usepackage{booktabs}%
\usepackage{algorithm}%
\usepackage{algorithmicx}%
\usepackage{algpseudocode}%
\usepackage{listings}%

\newboolean{showcomments}
\setboolean{showcomments}{true}
\ifthenelse{\boolean{showcomments}}
{\setboolean{showcomments}{true}
\ifthenelse{\boolean{showcomments}}
{\newcommand{\nb}[3]{
		{\colorbox{#2}{\bfseries\sffamily\scriptsize\textcolor{white}{#1}}}
		{\textcolor{#2}{\sf\small$\blacktriangleright$\textit{#3}$\blacktriangleleft$}}}
	}
{\newcommand{\nb}[3]{}
	
}

}

\begin{document}

\title[Machine Learning Reveals Impact]{Machine Learning Reveals Large-scale Impact of Posidonia Oceanica on Mediterranean Sea Water}

\author[1]{Celio Trois}
\author[1]{Luciana Didonet Del Fabro}
\author[1]{Vladimir A. Baulin}

\affil*[1]{\orgdiv{Departament de Química Fisica i Inorganica}, \orgname{Universitat Rovira i Virgili}, \orgaddress{C/Marcelli Domingo}, \city{Tarragona}, \postcode{43007}, \country{Spain}}

\abstract{\textit{Posidonia oceanica} is a protected endemic seagrass of Mediterranean sea that fosters biodiversity, stores carbon, releases oxygen, and provides habitat to numerous sea organisms. Leveraging augmented research, we collected a comprehensive dataset of 174 features compiled from diverse data sources. Through machine learning analysis, we discovered the existence of a robust correlation between the exact location of \textit{P. oceanica} and water biogeochemical properties. The model's feature importance, showed that carbon-related variables as net biomass production and downward surface mass flux of carbon dioxide have their values altered in the areas with \textit{P. oceanica}, which in turn can be used for indirect location of \textit{P. oceanica} meadows. The study provides the evidence of the plant's ability to exert a global impact on the environment and underscores the crucial role of this plant in sea ecosystems, emphasizing the need for its conservation and management.}


\maketitle

\section{Introduction}
\label{sec:introduction}

\textit{Posidonia oceanica} is one of the most dominant seagrass species in the Mediterranean Sea~\cite{ruiz_atlas_2015}. It provides essential habitats for diverse marine animals including fish, octopus, squid, and other animals~\cite{otero_m_m_numa_c_a5535_nodate}. Only along the coast of Almeria in Spain, there have been almost 1000 species of plants and animals identified in \textit{P. oceanica} meadows; similarly, on the Murcian coast, over 150 species of mollusks have been described, and the leaves of \textit{P. oceanica} have been associated with the presence of more than 400 different species of seaweed and several thousand animal species. This remarkable abundance of biodiversity makes this underwater habitat a truly unique and ecologically rich. Furthermore, the \textit{P. oceanica} acts as a natural filter, enhancing water clarity and nutrient cycling, while also mitigating the impacts of coastal erosion by stabilizing sediment~\cite{campagne_seagrass_2015}.

\textit{P. oceanica}, with its wide distribution and well-studied biology and ecology, has been used as an indicator organism~\cite{lopez_y_royo_seagrass_2011} for assessing environmental quality. Its response to anthropogenic disturbances has allowed researchers to establish correlations between certain structural and functional variables and the quality of the surrounding waters. However, gaining a comprehensive understanding of the complex relationship between \textit{P. oceanica} and its surrounding aquatic environment continues to pose a challenge. There are several crucial biological, chemical, and physical variables that need to be taken into account when predicting the presence and well-being of \textit{P. oceanica} meadows~\cite{romero_multivariate_2007}. These variables, on one hand can offer insights into the environmental conditions and nutrient availability that directly impact the growth and survival of this seagrass species~\cite{cocozza2011chemical}, and on other hand, they can be used as indicators for the presence and health of the \textit{P. oceanica} meadows~\cite{lopez_y_royo_seagrass_2011}. Identification of these variables aims to gain a deeper understanding of the biogeochemical conditions and their interrelations that contribute to the creation of a favorable environment for the thriving of \textit{P. oceanica} to ensure the sustainability of the seagrass ecosystem \textit{a-priori}, without the need of collecting field data that requires significant resources and expenses~\cite{effrosynidis2019seagrass}. 

In this sense, \cite{romero1994belowground} discusses the accumulation of carbon, nitrogen, and phosphorus in the below ground parts of \textit{P. oceanica}, including roots, leaf bases, rhizomes, and organic matter in the sediment. They highlight its role as a significant biogeochemical sink in the Mediterranean sea. Also, a study reported by~\cite{champenois2021net} estimates the gross primary production, community respiration, and net community production in a \textit{P. oceanica} seagrass meadow, computed from O$_2$ measurements. They concluded that the meadow was net autotrophic, in agreement with oxygen oversaturation (104\% at annual scale) and also acts as a CO$_2$ sink.

There are several works studying the effects of seagrasses on biochemical variables for specific locations; however, some authors use environmental data to feed Machine Learning (ML) models aiming to create Species Distribution Models or Habitat Suitability Models. \cite{bertelli2022use} reviewed the current knowledge on methodological approaches used to model and map habitat suitability for coastal ecosystems. They collated 75 publications, of which 35 included seagrasses, finding that the most commonly used variables for predicting seagrasses were bathymetry (74\%), salinity (57\%), light availability (51\%), and temperature (51\%). Also, by merging datasets about seagrass presence and other external environmental variables, \cite{effrosynidis2018seagrass} created a dataset aiming to determine the most appropriate variables affecting the distribution of seagrasses. Their experiments yield an accuracy up to 93\% in detecting seagrass presence and from their variable, they concluded that Cymodocea and Posidonia favor the low chlorophyll levels, followed by distance-to-coast, and bathymetry data.

This study aims to demonstrate the influence of seagrass on global water conditions, correlating biological, physical, and chemical variables with the precise positioning of \textit{P. oceanica} meadows. By utilizing an augmented research approach that facilitates the re-purposing, adapting, and reusing of multiple datasets, we created a labeled dataset with more than 6000 locations, pinpointing the existence or absence of \textit{P. oceanica}. This dataset combines 12 biogeochemical variables such as oxygen, carbon flux and pressure, biomass production, phosphate, ammonium, and nitrate, with physical variables depth, water temperature, salinity, and transparency. The monthly average values of these variables were collected over a period of one year from different open sources, including historical water data, oceanographic data, and satellite imagery, resulting in a dataset with 174 parameters. 

The dataset parameters were used as features to train ML models, aiming to identify the exact location of \textit{P. oceanica} meadows. However, due to strong correlations among variables and to enhance computational efficiency, dimensionality reduction techniques were applied to eliminate the most highly correlated features. It was observed that a removal threshold of 0.8 allowed the classifiers to maintain the same level of accuracy and precision in locating the meadows. This value was then used to investigate which features were most important in the classification process. 

The predictive capacity of the model enables a thorough investigation of the impact of \textit {P. oceanica} on biochemical concentrations in water. This analysis provides valuable insights into the presence of \textit {P. oceanica} and contributes to the development of evidence-based conservation strategies for the Mediterranean marine ecosystem.


\section{Results}
\label{sec:results}

Our objective is twofold: first, to establish a correlation between the geographic locations of \textit{P. oceanica} and the biogeochemical and physical variables of seawater; and second, investigate, through statistical analyzes inherent to ML models and by inspecting the correlations among biogeochemical variables, unveiling which of them are affected by the existence of \textit{P. oceanica}. 

\subsection{Key Biogeochemical Variables}
\label{sec:eval_biogeochem}

First, we focus on the exploration of predictive models using only biogeochemical variables extracted from the Mediterranean Sea Biogeochemical Reanalysis (MSBR)~\cite{cossarini2021high}. The dataset contains the monthly average values of all 12 variables available in the MSBR. Data was collected for one year, ranging from July 2020 to June 2021, as this was the most recent data available in the MSBR. For further information on how the dataset was created, refer to Section~\ref{sec:biogeochemvar}. 

Throughout this text we use the notation ``\texttt{variable}'' when referring to biochemical compounds, e.g. oxygen (\texttt{O$_2$}), phosphate (\texttt{PO$_4$}), net primary production of biomass (\texttt{nppv}), or physical variables, such as salinity (\texttt{so}), temperature (\texttt{thetao}), etc. For a complete list of variables, see Table~\ref{tab:dataset_variables} and Table~\ref{tab:dataset_variables2}. In the same way, we use the term ``\texttt{feature}'' when we make reference to the values of a variable in a given month, for example, the values of oxygen in the month of January 2021 is represented by the feature [\texttt{O$_2$ (Jan.2021)}].

\begin{figure*}[ht!]
   \centering
        \includegraphics[width=.85\textwidth]{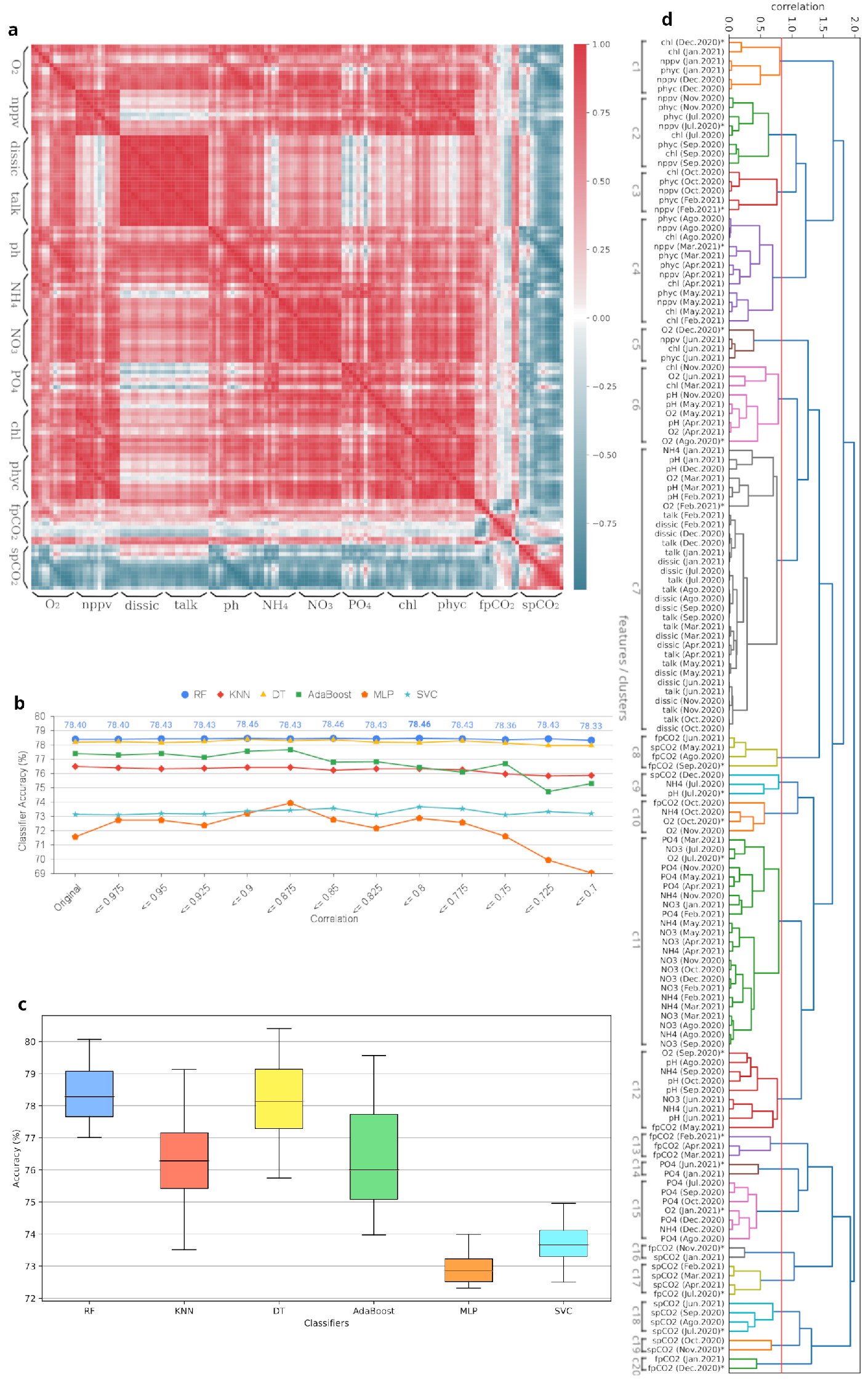}
        \caption{\textbf{Machine learning models performance and dimensionality reduction. a}, Heat map showing the features correlations. \textbf{b}, Classification accuracy on reducing the correlated features. \textbf{c}, Classifiers accuracy on features with correlation $<=$ 0.8. \textbf{d}, Hierarchical clustering dendrogram.}
         \label{fig:fig1}
\end{figure*}

Figure~\ref{fig:fig1}.a shows the correlation among the biogeochemical features, where the first 12 rows/columns refer to the \texttt{O$_2$} values ranging from July 2020 to June 2021, the next groups of 12 correspond to the other biogeochemical variables in the dataset (Table~\ref{tab:dataset_variables}). As can be seen, there exist many positive correlations values (dark red) and negative correlations (dark blue), meaning that they are redundant, irrelevant, or noisy. For example, the variables \texttt{dissic} and \texttt{talk}, for all months are highly correlated, as they appear as a red square in the upper-left corner. 

As many features exhibit a strong correlation and thus removing them can improve the dataset quality~\cite{ladha2011feature}, we have conducted an investigation into dimensionality reduction. This involves filtering out the correlated features and retaining only those with the most significant impact on the classification process. Initially, we analyzed the correlations among all features using the Pearson correlation coefficient~\cite{benesty2008importance}; next, we applied dimensionality reduction to improve interpretability and computational efficiency. 

We applied the methodology detailed in Section~\ref{sec:model_evaluation} to assess the classifiers accuracy upon removing the correlated variables. As depicted in Figure~\ref{fig:fig1}.b, certain classifiers experienced a noticeable decrease in accuracy when the feature correlation removal threshold is lower than 0.8. That is, significant features were pruned, leading us to opt for a reduction threshold of 0.8.

Figure~\ref{fig:fig1}.c highlights the accuracy for all classifiers, considering the correlation between the variables $\leq 0.8$. As can be seen, the best results were achieved by the RF classifier, with an average prediction accuracy of 78.46\%, followed by DT with 78.16\%, AdaBoost with 76.43\%, and K-Nearest Neighbors (KNN) with 76.33\%; the other classifiers demonstrated an accuracy lower then 76\%. 

The selection criteria for correlation reduction of 0.8 pruned the dataset to 20 features. The dendrogram shown in Figure~\ref{fig:fig1}.d represents a hierarchical tree with the relationships among the features. In the bottom, all features are listed as leaf nodes, and going up in the y axis, branches connect these nodes based on their correlation, generating clusters of correlated features. The red  horizontal line represents the chosen threshold of 0.8, and bellow this line, it is possible to see, in different colors, the 20 remained features and their correlations, enumerated as \texttt{c1} to \texttt{c20}. For example, the gray cluster \texttt{c7} corresponds to the red square in the upper corner of Figure~\ref{fig:fig1}.a, representing all features of the variables \texttt{nppv}, \texttt{dissic}, and \texttt{talk}. In this case, the feature that represents this cluster is \texttt{O$_2$ (Feb.2021)}, highlighted in the dendrogram with an asterisk next to it. Table~\ref{tab:clusters} presents the features that represent each cluster  and their clustered features. 

\begin{table}[ht!]
\caption{Correlated features.}
\label{tab:clusters}
\begin{tabular}{cll}
\toprule
\textbf{cluster} & \textbf{Feature} & \textbf{Clustered features} \\ 
\midrule
c1 & chl (Dec.2020) & \begin{tabular}[c]{@{}l@{}}chl (Jan.2021)\\ nppv (Dec.2020, Jan.2021)\\ phyc (Dec.2020, Jan.2021)\end{tabular} \\
c2 & nppv (Jul.2020) & \begin{tabular}[c]{@{}l@{}}chl (Jul.2020)\\ nppv (Nov.2020, Sep.2020)\\ phyc (Nov.2020, Jul.2020, Sep.2020)\end{tabular} \\
c3 & nppv (Feb.2021) & \begin{tabular}[c]{@{}l@{}}chl (Oct.2020)\\ nppv (Oct.2020)\\ phyc (Oct.2020, Feb.2021)\end{tabular} \\
c4 & nppv (Mar.2021) & \begin{tabular}[c]{@{}l@{}}chl (Ago.2020, Apr.2021, May.2021, Feb.2021)\\ nppv (Ago.2020, Apr.2021, May.2021)\\ phyc (Ago.2020, Mar.2021, Apr.2021, May.2021)\end{tabular} \\
c5 & O$_2$ (Dec.2020) & \begin{tabular}[c]{@{}l@{}}chl (Jun.2021)\\ nppv (Jun.2021)\\ phyc (Jun.2021)\end{tabular} \\
c6 & O$_2$ (Ago.2020) & \begin{tabular}[c]{@{}l@{}}chl (Nov.2020, Mar.2021)\\ O$_2$ (Jun.2021, May.2021, Apr.2021)\\ pH (Nov.2020, May.2021, Apr.2021)\end{tabular} \\
c7 & O$_2$ (Feb.2021) & \begin{tabular}[c]{@{}l@{}}dissic (all months)\\ NH$_4$ (Jan.2021)\\ O$_2$ (Mar.2021)\\ pH (Dec.2020, Jan.2021, Feb. 2021, Mar.2021, )\\ talk (all months)\end{tabular} \\
c8 & fpCO$_2$ (Sep.2020) & \begin{tabular}[c]{@{}l@{}}fpCO$_2$ (Jun.2021, Ago.2020)\\ spCO$_2$ (May.2021)\end{tabular} \\
c9 & pH (Jul.2020) & \begin{tabular}[c]{@{}l@{}}spCO$_2$ (Dec.2020)\\ NH$_4$ (Jul.2020)\end{tabular} \\
c10 & O$_2$ (Oct.2020) & \begin{tabular}[c]{@{}l@{}}fpCO$_2$ (Oct.2020)\\ NH$_4$ (Oct.2020)\\ O$_2$ (Nov.2020)\end{tabular} \\
c11 & O$_2$ (Jul.2020) & \begin{tabular}[c]{@{}l@{}}NH$_4$ (Nov.2020, Feb.2021, Mar.2021, Apr.2021, May.2021, \\ Ago.2020)\\ NO$_3$ (all months except Jun.2021)\\ PO$_4$ (Nov.2020, Feb.2021, Mar.2021, Apr.2021, May.2021)\end{tabular} \\
c12 & O$_2$ (Sep.2020) & \begin{tabular}[c]{@{}l@{}}fpCO$_2$ (May.2021)NH$_4$ (Sep.2020, Jun.2021)\\ NO$_3$ (Jun.2021)\\ pH (Ago.2020, Sep.2020, Oct.2020, Jun.2021)\end{tabular} \\
c13 & fpCO$_2$ (Feb.2021) & fpCO$_2$ (Mar.2021, Apr.2021) \\
c14 & PO$_4$ (Jun.2021) & PO$_4$ (Jan.2021) \\
c15 & O$_2$ (Jan.2021) & \begin{tabular}[c]{@{}l@{}}NH$_4$ (Ago.2020, Dec.2020)\\ PO$_4$ (Jul.2020, Sep.2020, Oct.2020, Dec.2020)\end{tabular} \\
c16 & fpCO$_2$ (Nov.2020) & spCO$_2$ (Jan.2021) \\
c17 & fpCO$_2$ (Jul.2020) & spCO$_2$ (Feb.2021, Mar.2021, Apr.2021) \\
c18 & spCO$_2$ (Jul.2020) & spCO$_2$ (Ago.2020, Sep.2020, Jun.2021) \\
c19 & spCO$_2$ (Nov.2020) & spCO$_2$ (Oct.2020) \\
c20 & fpCO$_2$ (Dec.2020) & fpCO$_2$ (Jan.2021) \\ 
\bottomrule
\end{tabular}
\end{table}

Despite achieving an accuracy rate of nearly 80\%, upon analyzing the model's predicted values, we identified a problem known as the ``pixelation''. This issue arises due to the spatial resolution of the biogeochemical variables in the dataset, which is $ 1/24^{\circ}$ (approximately 4 km). Consequently, all data points within the ``pixel'' have the same value. As a result, the model assigns the same label to all points within a pixel. To address this issue, we aggregated additional variables, with different resolutions, as described in the next section.

\subsection{Detecting P. oceanica with larger set of indirect indicators}

Although the model reports an average accuracy near 80\% using only biogeochemical variables, a limitation was noticed regarding the spatial resolution. To alleviate this issue, we repeated the same steps described in Section~\ref{sec:eval_biogeochem}, but expanding the dataset with additional variables pointed out by~\cite{bertelli2022use} as important to identify the locations of \textit{P. oceanica}, which are depth, water temperature, salinity, and transparency.

The variables salinity and temperature were gathered from the Global Ocean Analysis and Forecast (GOAF)~\cite{lellouche2019quality} also considering monthly average values for an entire year. The \texttt{depth} was collected from the European Marine Observation and Data Network (EMODnet) Digital Bathymetry and the water transparency was gathered from the Joint Research Centre (JRC) data. Further information on the expanded dataset is presented in Section~\ref{sec:othervariables}. 

After expanding the dataset, we performed the same dimensionality reduction tests and arrived at the same correlation cutoff of 0.8. It resulted in a set of 27 features, appending to the previous dataset the \texttt{depth}, the water transparency (\texttt{transp}), salinity (\texttt{so}) of February, March, and April, and water temperature (\texttt{thetao}) of January, June, and October.

\begin{figure*}[t]
         \centering
    \includegraphics[width=.9\textwidth]{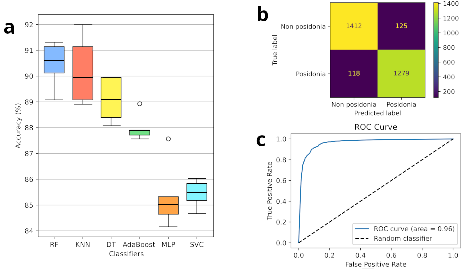}
         \caption{\textbf{Machine learning models performance evaluation on biogeochemical, depth, water temperature, salinity, and transparency, considering variables correlation~$\le  0.8$. a}, Classifiers accuracy. \textbf{b}, Confusion Matrix. \textbf{c}, Receiver Operating Characteristic (ROC) curve.}
         \label{fig:fig2}
\end{figure*}

We repeated the same accuracy test presented in the previous section on the expanded dataset. Figure~\ref{fig:fig2}.a shows a boxplot with the accuracy of all classifiers when trained with these features. The RF model stands out with an average accuracy of 90.45\%, indicating its proficiency in discerning and classifying the \textit{P. oceanica} locations. Following closely is the KNN with an accuracy of 90.21\%. DT model exhibits an accuracy of 89.09\%, while the AdaBoost achieves an accuracy of 88.0\%. 

The confusion matrix provides a detailed breakdown of the model's performance by presenting the number of true positive (TP), true negative (TN), false positive (FP), and false negative (FN) predictions. Figure~\ref{fig:fig2}.b shows the confusion matrix; the top-left cell (1,412) corresponds to the number of true negatives, indicating instances correctly classified as \texttt{Non Posidonia}. The top-right cell (125) represents false positives, signifying instances incorrectly predicted as \texttt{Posidonia} when they were actually \texttt{Non Posidonia}. The bottom-left cell (118) corresponds to false negatives, indicating instances misclassified as \texttt{Non Posidonia} when they were actually \texttt{Posidonia}. Lastly, the bottom-right cell (1,279) represents true positives, denoting instances correctly classified as \texttt{Non Posidonia}. 

As can be seen in Figure~\ref{fig:fig2}.c, the Receiver Operating Characteristic (ROC) curve is plotted closely to the top-left corner, indicating a combination of high sensitivity and low false positive rate. The steepness of the curve implies a rapid transition between sensitivity and specificity, allowing for an effective trade-off in classification thresholds. Additionally, its performance suggests suitability for tasks where precision in positive predictions and minimizing misclassifications are paramount. It presents an Area Under the Curve (AUC) of 0.96, reinforcing the potential of the model for reliable real-world applications.

Table~\ref{tab:rf_classif_repa} presents the classification report of RF model. The precision values for \texttt{Non Posidonia} (0.9236) and \texttt{Posidonia} (0.9183) indicate the accuracy of the model in correctly identifying instances of each class. The recall values, measuring the model's ability to capture all relevant instances, are 0.9206 for \texttt{Non Posidonia} and 0.9162 for \texttt{Posidonia}. The F1-scores, which balance precision and recall, are 0.9221 for \texttt{Non Posidonia} and 0.9146 for \texttt{Posidonia}. The overall accuracy of 0.9185 signifies the proportion of correctly classified instances. The macro average, considering unweighted class contributions, is 0.9183, and the weighted average, accounting for class imbalances, is 0.9186. These metrics collectively demonstrate the performance of the model in delineating between \texttt{Posidonia} and \texttt{Non Posidonia} classes, demonstrating that the addition of new variables increased the prediction metrics.

\begin{table}[t]
\caption{Random Forest classification report using biogeochemical, bathymetry, water temperature, salinity, and transparency variables.}
\label{tab:rf_classif_repa}
\begin{tabular}{lcccc}
\toprule
     & Precision & Recall & \multicolumn{1}{c}{F1-score} & Support                  \\ 
\midrule
\texttt{Non Posidonia}  & 0.9236      & 0.9206   & 0.9221                         & 1537                     \\
\texttt{Posidonia}      & 0.9183      & 0.9162   & 0.9146                         & 1397                     \\
Accuracy   -   &      -     &        & 0.9185                         & 2934 \\
macro avg.     & 0.9183      & 0.9184   & 0.9184                         & 2934                     \\
Weighted avg.  & 0.9186      & 0.9185   & 0.9185                         & 2934                     \\ 
\bottomrule
\end{tabular}%
\end{table}

\subsection{Visualizing the Predicted Locations}

\begin{figure}[t]
         \centering
         \includegraphics[width=\textwidth]{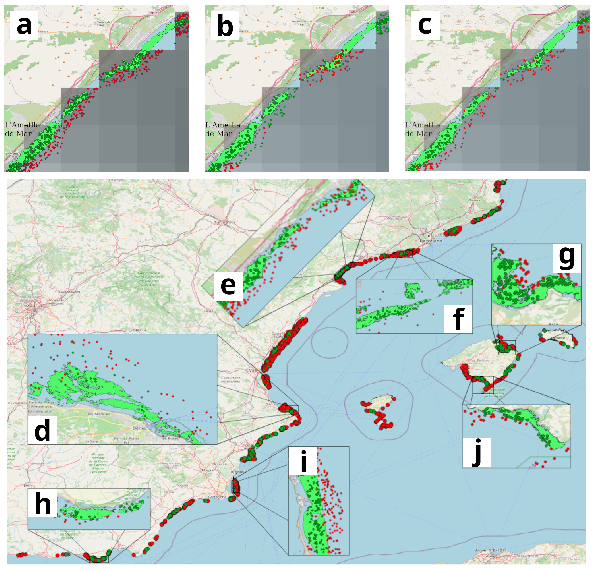}
         \caption{\textbf{Map of studied area, highlighting predicted values. a,} Generated data points near L’Ametlla de Mar, Catalonia. \textbf{b}, Pixel problem on predictions with biogeochemical variables. \textbf{c}, Predictions using biogeochemical, bathymetry, water temperature, salinity, and transparency variables. \textbf{d}, Zoom in on the P. oceanica grasslands of the city of Denia, Valencian community. \textbf{e}, L’Ametlla de Mar. \textbf{f}, Vilanova i la Geltrú, Catalonia. \textbf{g}, Alcudia Bay, north coast of Mallorca island. \textbf{h}, Almerimar, Andalucia. \textbf{i}, La Manga, Murcia. \textbf{j}, Colonia de San Jordi, southern part of Mallorca, Balearic Islands.}
         \label{fig:fig3}
\end{figure}

The Figure~\ref{fig:fig3}.a shows the data points in L’Ametlla de Mar, a municipality within the region of Baix Ebre, situated in the south of Catalonia. The green areas represent the \textit{P. oceanica} meadows; the green points inside these areas are the selected random points for \texttt{Posidonia} locations, while red points represent \texttt{Non Posidonia}. Gray shading in Figure~\ref{fig:fig3}.a, Figure~\ref{fig:fig3}.b, and Figure~\ref{fig:fig3}.c corresponds to chlorophyll mass concentration data in sea water [\texttt{chl}], and it was highlighted in the images to facilitate understanding of the pixel problem.

Figure~\ref{fig:fig3}.b enlarges the
predictions in L’Ametlla de Mar when using only the biogeochemical variables. It is possible to see that all predictions inside a ``pixel'' (different shadows of gray) have the same value as \texttt{Posidonia}, in green or \texttt{Non Posidonia}, in red. Figure~\ref{fig:fig3}.c shows the predictions for the same area, but using biogeochemical, bathymetry, water temperature, salinity, and transparency. It can be seen the model improvement, as now it has only a few misclassified points, \textit{i.e.}, red points inside of the green area or green points outside. 

Finally, a general map plots all predicted points for the test dataset. From Figure~\ref{fig:fig3}.d to Figure~\ref{fig:fig3}.j, some areas with a large number of \textit{P. oceanica} were enlarged to allow viewing the predictions. The model presents excellent accuracy and precision in its predictions, making it possible to understand the affinity of the variables used in the prediction process with the \textit{P. oceanica}.

\subsection{Impact of P. oceanica on Sea Water}

This section presents an investigation into the importance of features that drive the model's predictive ability, allowing us to understand the influence of \textit{P. oceanica} on the subtle changes in biogeochemical determinants. Features importance are calculated as the mean and standard deviation of the impurity decay accumulation within each tree of the Random Forest classifier.

\begin{figure*}[t]
         \centering
    \includegraphics[width=\textwidth]{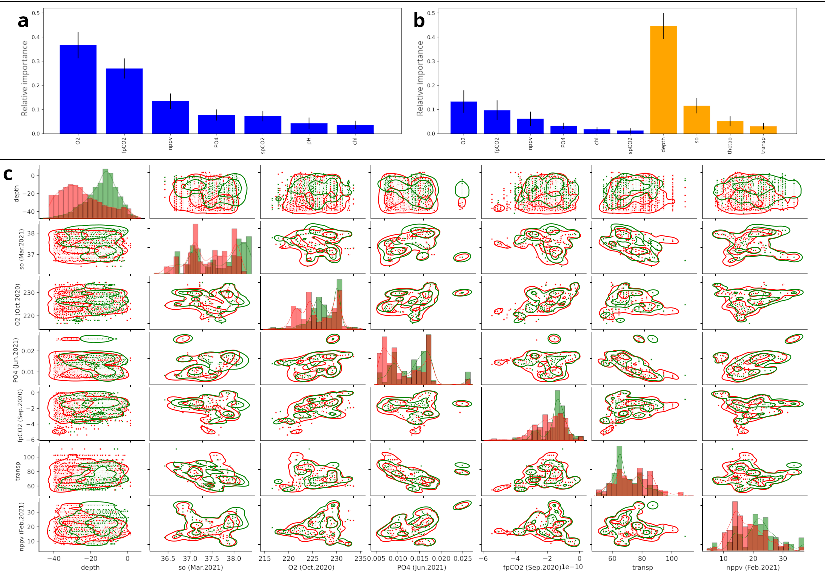}
         \caption{\textbf{Random Forest feature importance. a}, Using only biogeochemical data. \textbf{b}, Using biogeochemical, depth, water salinity, temperature, and transparency. \textbf{c}, KDE pairwise relationships; green refers to \texttt{Posidonia} data, while red indicates \texttt{Non Posidonia}.}
         \label{fig:fig4}
\end{figure*}

Considering only the biogeochemical variables, oxygen (\texttt{O$_2$}) was the first, with an importance of 0.367, calculated by summing the importances of \texttt{O$_2$} for the months of July, August, September, December of 2020, and January and February 2021. The second most important was the flux of carbon dioxide (\texttt{fpCO$_2$}), with an importance of 0.270, where the months of July, September, November, and December 2020 and February 2021 were used to calculate it. The net primary production of biomass (\texttt{nppv}) of July 2020, February and March 2021 was the third more important feature (0.134). Phosphate (\texttt{PO$_4$}) of June 2020 with 0.077 was the fourth more important, followed by the surface partial pressure of carbon dioxide in sea water (\texttt{spCO$_2$}) of July and November 2020 (0.072). The \texttt{pH}, and mass concentration of chlorophyll (\texttt{chl}) appear with a lower relative importance. Figure~\ref{fig:fig4}.a shows the importance when using only biogeochemical variables; the blue bars are the feature importance of the forest, along with their inter-trees variability represented by the error bars.

In the expanded dataset, \texttt{depth} had the highest importance, with 0.446, \texttt{O$_2$} appears as the second most important variable (0.133), followed by salinity (\texttt{so}) and \texttt{fpCO$_2$}, with 0.116 and 0.097, respectively. The \texttt{nppv} had an importance of 0.057, followed by sea water potential temperature (\texttt{thetao}) with 0.051. Phosphate, water transparency (\texttt{transp}), \texttt{chl}, and \texttt{spCO$_2$} obtained a relative importance less than 0.05. The Figure~\ref{fig:fig4}.b shows the importance, where biogeochemical data are plotted in blue and the other variables in orange. 

For a better understanding of how these features represent the classes (\texttt{Posidonia} and \texttt{Non Posidonia}), the Figure~\ref{fig:fig4}.c plots matrix with the pairwise relationships of the seven more important features, \texttt{depth}, \texttt{so} of March 2021, \texttt{O$_2$} of October 2020, \texttt{PO$_4$} of June 2021, \texttt{fpCO$_2$} of September 2020, \texttt{transp}, and \texttt{nppv} of February 2021. The secondary diagonal displays the histogram and the Kernel Density Estimation (KDE), showing the distribution of observations of these variables. The upper and lower (off-diagonal) triangles show the Bivariate KDE, employing contours to depict the density of points in the two-dimensional space. 

The \texttt{depth} is the only variable that visually presents clusters for the predicted classes. Despite the absence of clear separations between each pair of variables, the model, with its ability to consider multiple variables simultaneously was able to discern the intricate relationships. It highlights the versatility of Random Forests in capturing nuanced patterns, even when conventional grouping suggestions are elusive. The impossibility of perceiving isolated clusters for other features highlights the fact that individually they are insufficient for distinguishing meadow locations, but when analyzed together, they are effective in predicting the locations of \textit{P. oceanica}. 

\section{Discussion}

Our analysis has identified the most crucial biogeochemical variables associated with \textit{P. oceanica}, allowing to determine its location with an accuracy and precision of almost 80\%. However, we recognized that the performance was constrained by the coarse resolution of the data. To address this limitation, we incorporated additional variables, measured on a different scale and commonly used in the literature for predicting this seagrass (depth, water salinity, temperature, and transparency). As a result, we achieved an accuracy and precision exceeding 90\%. In this section, we delve into a discussion of these variables to demonstrate their causal or consequential relationship with \textit{P. oceanica} based on existing literature.

First, let's consider non-biochemical variables that influence the presence of seagrass; it includes physical variables that regulate its physiological activity, natural phenomena, and anthropogenic pressures~\cite{telesca_seagrass_2015}. Starting with \texttt{depth}, the lower limit of \textit{P. oceanica} meadows typically ranges from 25 to 40 meters, and can be even deeper in areas with exceptionally clear waters, such as those found in the Balearic Islands~\cite{leriche2004one}. The upper limit of the \textit{P. oceanica} meadows is determined by hydrodynamics, specifically the wave regime. Wave action can damage the plants and disturb the seabed, making it difficult for the plants to survive. In meadows located in sheltered coves where wave action is minimal, the plants can grow to the surface, as observed in the northern region of Menorca. However, as a general rule, the upper limit is usually found at a depth of 3 to 5 meters, where the wave forces are not strong enough to dislodge the plants from the seabed. The majority of meadows are typically situated at depths ranging from 5 to 20 meters~\cite{ruiz_atlas_2015}, which is consistent with the depth data histogram presented in Figure~\ref{fig:fig4}.c. The adjacent histogram in the figure, along the secondary diagonal, illustrates salinity, with the recorded values aligning with \cite{telesca_seagrass_2015}, who reported that \textit{P. oceanica} is a stenohaline species, thriving within a salinity range of 36.5 to 39.5 ppt.

To achieve an optimal development, \textit{P. oceanica} requires a favorable environment with transparent waters. The level of water transparency directly impacts the plant's ability to harness the sun's rays for energy through photosynthesis. Therefore, light plays a crucial role in regulating the presence of \textit{P. oceanica}. Upon observing the transparency histogram in Figure~\ref{fig:fig4}.c, it is evident that areas with \textit {P. oceanica} have lower values (between 50 and 70), indicating higher transparency~\cite{martin2023deep}, as it is based on the diffuse attenuation coefficient \texttt{Kd490}.

When considering the biogeochemical variables, it is important to understand that the functioning of a \textit {P. oceanica} meadow relies on photosynthesis taking place in the chloroplasts of the leaves. Here, chlorophyll captures light and converts it into chemical energy~\cite{ruiz2012praderas}. This energy, together with nutrients (nitrogen and phosphorus) from water and sediments, forms organic matter, sustaining the primary production of the ecosystem. Higher levels of PO$_4$ are present in both the water column and sediment pore water at night on a daily basis, particularly in warm seasons, due to increased decomposition and sediment release under low-oxygen conditions. These findings are depicted in the PO$_4$ histogram in Figure~\ref{fig:fig4}.c. It is important to note that nitrogen-containing variables (NO$_3$ and NH$_4$) did not appear as important features, that is explained because they were filtered out during the dimensionality reduction due to their high correlation with phosphate~\cite{touchette2000review}, as can be seen in the clusters \texttt{c11}, \texttt{c14}, and \texttt{c15} in the Figure~\ref{fig:fig1}.d.

\textit{P. oceanica} is highly productive compared to other seagrass systems. The meadows generate between 4 and 20 liters of oxygen per square meter per day, constituting one of the most important sources of oxygenation in the Mediterranean~\cite{duarte1999seagrass}. The oxygen histogram (Figure~\ref{fig:fig4}.c) shows that the largest number of points sampled with \textit{P. oceanica} present a higher value, in relation to the points without this seagrass, that is, on a large scale it is possible to perceive the effect of \textit{P. oceanica} on generating oxygen in the sea. Part of this oxygen is released into the Earth's atmosphere during periods of maximum productivity capturing carbon dioxide (CO$_2$). Seagrass meadows are mainly autotrophic, their productivity exceeds their carbon needs, demonstrating a high capacity to capture CO$_2$, representing 24.3\% of global marine carbon sequestration, they are also referred to as blue carbon habitats~\cite{duarte2013role,macreadie2019future}.

As can be seen in Figure~\ref{fig:fig1}.d, there is a strong correlation between surface downward mass flux of carbon dioxide (\texttt{fpCO$_2$}) and surface partial pressure of carbon dioxide (\texttt{spCO$_2$}), which are grouped into seven clusters (\texttt{c8}, \texttt{c13}, \texttt{c16}, \texttt{c17}, \texttt{c18}, \texttt{c19}, \texttt{c20}), as the exchange of CO$_2$ between the ocean and the atmosphere is controlled by the air–sea difference in partial pressure of CO$_2$ at the surface and of the efficiency of the transfer processes~\cite{RUTGERSSON2008381}.

Sea water alkalinity (\texttt{talk}) and dissolved inorganic carbon (\texttt{dissic}) are two parameters in ocean chemistry that are closely related to each other. The process occurs when CO$_2$ from the atmosphere dissolves in seawater, leading to a series of chemical reactions that result in the formation of carbonic acid (H$_2$CO$_3$). This weak acid dissociates into bicarbonate (HCO$_3^-$) and carbonate ions (CO$_3^{2-}$), both of which release hydrogen ions (H$^+$) that lower the pH of seawater~\cite{cox2015effects}. This correlation is clear in the cluster \texttt{c7}, on Figure~\ref{fig:fig1}.d, which groups all \texttt{talk} and \texttt{dissic} features. 

Biomass production (\texttt{nppv}) fluctuates seasonally in response to changes in light and temperature. Leaf growth reaches its peak during spring and summer when there is ample light and heat, but slows down in winter. Seagrasses store carbon from summer photosynthesis as starch in the rhizome, utilizing it in autumn and winter when nutrient availability is higher~\cite{ruiz2012praderas}. During the onset of the new growing season in autumn, elevated cultivation temperatures have been observed to enhance photosynthetic rates~\cite{pirc1986seasonal}, explaining the histogram of \texttt{nppv} (Figure~\ref{fig:fig4}.c), where the production of biomass, is higher in areas with \textit{P. oceanica}. Biomass, phytoplankton, and chlorophyll are closely correlated in seawater due to the fundamental role of phytoplankton in chlorophyll production and their contribution to overall biomass~\cite{gobert2002posidonia}; these correlations are clear in the clusters \texttt{c1}, \texttt{c2}, \texttt{c3}, \texttt{c4}, and \texttt{c5}, shown in Figure~\ref{fig:fig1}.d. 

This section showed that our machine learning statistical analyzes are aligned with current literature, reinforcing that the correlations identified between biogeochemical variables are linked to the presence of Posidonia oceanica. 


\section{Methods}
\label{sec:methods}

This section describes the materials and methods used to obtain the results presented in this article. First, we present the process of creating the dataset, then the tools and methods used in cleaning and processing the data and, finally, the development and evaluation metrics of the model are detailed.

\subsection{Dataset Creation}
\label{sec:dataset}

For creating the dataset, we have first researched information where the \textit{P. oceanica} is located. Then, we algorithmically generated random points in and around these locations, pinpointing where there is and there is not \textit{P. oceanica}. Next, we gathered biochemical and physical data for each generated point, discarding the points where there was no data availability. 

\subsubsection{Locations}

For mapping the locations of \textit{P. oceanica} meadows, we have used data from the five Spanish autonomous communities which are bathed by the Mediterranean Sea, namely Catalonia, Valencia, Murcia, Andalusia, and Balearic Islands. 


The coast of the Catalan Community extends in a general NE-SW direction for 699 km according to the National Institute of Statistics, between the parallels 42$^o$ 51.83' and 40$^o$ 31.46' N. The climate is typically Mediterranean, with an average annual air temperature of 17$^o$C, warm summers and mild winters 23$^o$C and 12$^o$C on average, respectively. The precipitation is relatively low. The temperature of marine surface waters ranges, in open waters, between 12-13$^o$C in winter and 22-23$^o$C (North end of the coast) and 24-25$^o$C (Southern end) at the end of summer. These values can be exceeded in shallow semi-confined waters, such as the Alfacs bay (Delta of Ebro), where minimum temperatures of 7$^o$C and maximum temperatures of 30$^o$C have been observed.

\textit{P. oceanica} meadows are of two types, according to coastal morphology. On the one hand, on sandy and rectilinear coasts, meadows appear as longitudinal, parallel formations at a certain distance from the coast (from hundreds of meters to more than a mile). The deep limit is marked, in most cases, by the penetration of light. This deep limit ranges between 17 and 20 meters, sporadically up to 25. The Department of Climate Action, Food and Rural of Catalonia provides data containing the delimitation of the grasslands or meadows of marine phanerogams\footnote{Available at: \url{https://t.ly/879lk}}
known along the coast of its community.


The Valencian Community is located on the eastern coast of the Iberian Peninsula. Geographically, it extends from the Sénia River to Pilar de la Horadada, beyond the mouth of the Segura River, with a coastline 644 km long. Throughout the year, the weather is dry and, in general, the temperature varies from 6$^o$C to 30$^o$C. The average water temperature goes through extreme seasonal variations throughout the year. In summer the average water temperature is above 23$^o$C while in winter the water temperature is below 16$^o$C.

\textit{P. oceanica} has an uneven distribution on the Valencian coast: it is widely distributed on the Alicante coast, while on the Valencian and Castellón coasts it has a very reduced distribution. The Department of Agriculture, Rural Development, Climate Emergency and Ecological Transition makes available the mapping and cataloging of seagrass meadows in this community\footnote{Available at: \url{https://t.ly/1g4D4}}.



The Murcian Community has 224 km of coastline is part of the Alicante-Almería coastal axis, which forms the only area with a semi-arid Mediterranean climate in the national territory. The average annual sunshine is very high (2,800-3,000 hours) and the average annual rainfall (151 mm in Cabo Tiñoso) is one of the lowest in Europe. The temperature of the Mediterranean coastal waters on the Murcian coast varies between 29$^o$C in summer and 12$^o$C in winter.

The infra-littoral Mediterranean environments are dominated by the meadows of the species \textit{P. oceanica} to a maximum depth of between 25 and 30 meters, although it reaches up to 34 m off Grosa Island. It colonizes both sedimentary and rocky bottoms, in environments with fairly stable salinity (37-38 UPS), and is not found in hyper saline environments such as the Mar Menor, due to its inability to tolerate salinities greater than 38.5 UPS~\cite{RuízMarínGuiraoSandovalGil}. The locations of \textit{P. oceanica} are available in the Spatial Data Infrastructure of Spain, through
a Web Map Service\footnote{Available at: \url{https://shorturl.at/HQV06}}. 


In Andalusia, the temperate Mediterranean climate predominates, characterized by mild winters with irregular rainfall, and dry, hot and sunny summers, which accentuate its characteristics as you move from the coast to the interior of the region. It has an average annual temperature of approximately 18$^o$C, and more than 300 days of sunshine a year, with January being the coldest month and August the hottest. Andalusia is one of the warmest regions in Spain, with an average daily maximum temperature of 24$^o$C. In 6 months the average temperatures exceed 25$^o$C and in winter the minimum temperature reaches 6$^o$C. The water temperature, on average, varies between 15.5$^o$C and 23.8$^o$C.

On the Andalusian coast, \textit{P. oceanica} forms meadows on the infralittoral floor, from very shallow waters to a variable depth that depends on the transparency of the water, both on rocky and sandy substrates. For example, on the Almeria coast it can reach up to 30 meters deep and form very dense meadows. The Consejería de Sostenibilidad, Medio Ambiente y Economía Azul provides a map with the location of the Praderas de fanerógamas marinas\footnote{Available at: \url{https://shorturl.at/amAEO}}.



The Balearic Islands have privileged geographical and climatological conditions. Due to their location in the center of the western Mediterranean, they are the Mediterranean islands furthest from the coast, in a position equidistant between Africa, France, and Sardinia, and 170 km from the coast of eastern Spain. This situation conditions its climate, a sub-humid Mediterranean type, with an average annual rainfall between 350 and 650 mm, with high interannual variations, and mild average temperatures throughout the year, which vary between 5$^o$C and 31$^o$C depending on the time of year. The water temperature in the Balearic Islands varies throughout the year due to natural factors and climatic phenomena. During the winter months, the island experiences temperatures around 15$^o$C to 17$^o$C. In summer, sea temperatures can exceed 26$^o$C.

\textit{P. oceanica} occurs in infralittoral seabeds, from very shallow seabeds of 1 m depth to a maximum depth recorded in the Balearic Islands of 43 meters~\cite{marba2002effectiveness} on the island of Cabrera. In the Balearic Islands, \textit{P. oceanica} meadows develop both in soft sediments and on rock, and form the dominant climax ecosystem of these beds. In areas with a steeper platform, the upper limit of the meadow appears at greater depth due to hydrodynamism, and the meadow appears more fragmented and restricted to the rocky substrate. In this sense, the meadows of Menorca stand out, an island that, due to its geographical location, is subject to greater exposure to waves and storms. The Ministry of Agriculture, Fisheries and Natural Environment of the Balearic Islands, through the Balearic Institute of Nature, provides a detailed cartography of \textit{P. oceanica} locations\footnote{Available at: \url{https://shorturl.at/cjovL}}.


\subsubsection{Random Points Generation}

To create random points, we used the software QGIS\footnote{Available at: \url{https://qgis.org/}} version 3.22.16. We loaded all information of \textit{P. oceanica} locations in the QGIS and we used the \texttt{Random Points in Extent} in the whole area for generating the random points with a minimum distance between points of 50 meters. For getting the \textit{P. oceanica} points, we used the \texttt{Clip} tool for filtering the points inside \textit{Posidonia} meadows. The selection of Non \textit{P. oceanica} locations, we created a buffer area ranging from 50 meters to 5 kilometers away from the \textit{P. oceanica}. Duarte et al.~\cite{duarte1991seagrass} stated that \textit{P. oceanica} grows all along the coast forming extensive meadows from nearly the water surface to depths up to 40 meters. So, we reduced the buffer considering a maximum depth of 40 meters. Finally, for getting a similar number of points for location with and without \textit{P. oceanica}, we randomly selected the same number of points for both.

\subsubsection{Biogeochemical Variables}
\label{sec:biogeochemvar}

The Mediterranean Sea Biogeochemical Reanalysis~\cite{cossarini2021high} (MSBR) was developed within the Copernicus Marine Environment Monitoring Service framework~\cite{von2018copernicus}, the reanalysis integrates the Biogeochemical Flux Model with a variational data assimilation scheme, using NEMO–OceanVar and ECMWF ERA5 data~\cite{hersbach2020era5}. This dataset was validated through multiple observational sources, assessing 12 state variables, fluxes, and various metrics. According to the authors, it presents overall good skill in simulating basin-wide biogeochemical variables and variability, allowing diverse communities to explore spatial and temporal variability in Mediterranean biogeochemistry.

The original publication covered the 1999–2019 period with monthly means of 12 published and validated biogeochemical variables. However, a new version was released updating the horizontal resolution to $1/24^{\circ}$ and extension till June 2021\footnote{Available at: \url{https://t.ly/hYMqb}}.
The reanalysis assimilates surface data and integrates the European Marine Observation and Data Network (EMODnet) data~\cite{martin2019european} as initial conditions, in addition to considering World Ocean Atlas data at the Atlantic boundary, CO2 atmospheric observations, and yearly estimates of riverine nutrient inputs. Details on model explanation and evaluation are available in the original publication~\cite{cossarini2021high}.

For building the dataset, we selected all 12 variables available in the MSBR, getting the monthly average for 12 months, ranging from July 2020 until June 2021. We used these dates because they represent an entire year, considering the most recent data available in MSBR. The biogeochemical variables are summarized in the Table~\ref{tab:dataset_variables}.

\begin{table}[t]
\caption{Biogeochemical variables composing the dataset. Source MSBR~\cite{cossarini2021high}}
\label{tab:dataset_variables}
\begin{tabular}{llcc}
\toprule
\textbf{Var.}   & \textbf{Description}                                                        & \textbf{Unit}  \\ 
\midrule
\texttt{O$_2$}   & \begin{tabular}[c]{@{}l@{}}Mole concentration of dissolved molecular oxygen in sea water\end{tabular} & [$mmol/m^3$]  \\
\texttt{nppv}   & \begin{tabular}[c]{@{}l@{}}Net primary production of biomass expressed as carbon per unit \\  volume in sea water\end{tabular} & [$mg/m^3/day$] & \\
\texttt{dissic} & \begin{tabular}[c]{@{}l@{}}Mole concentration of dissolved inorganic carbon in sea water\end{tabular}                      & [$mol/m^3$]  &  \\
\texttt{talk}  & \begin{tabular}[c]{@{}l@{}}Sea water alkalinity expressed as mole equivalent\end{tabular}       & [$mol/m^3$] &  \\
\texttt{pH}    & Sea water pH reported on total scale.                                                              &              & \\
\texttt{NH$_4$}   & \begin{tabular}[c]{@{}l@{}}Mole concentration of ammonium in sea water\end{tabular}             & [$mol/m^3$] &  \\
\texttt{NO$_3$}   & \begin{tabular}[c]{@{}l@{}}Mole concentration of nitrate in sea water\end{tabular}                                                          & [$mol/m^3$]  & \\
\texttt{PO$_4$}   & \begin{tabular}[c]{@{}l@{}}Mole concentration of phosphate in sea water\end{tabular}            & [$mol/m^3$]  & \\
\texttt{chl}   & \begin{tabular}[c]{@{}l@{}}Mass concentration of chlorophyll in sea water\end{tabular}        & [$mg/m^3$]   & \\
\texttt{phyc}   & \begin{tabular}[c]{@{}l@{}}Mole concentration of phytoplankton expressed as carbon in \\ sea water\end{tabular}               & [$mol/m^3$]   & \\ 
\texttt{fpCO$_2$}  & \begin{tabular}[c]{@{}l@{}}Surface downward mass flux of carbon dioxide expressed as carbon\end{tabular}                   & [$kg/m^2/s$]  & \\
\texttt{spCO$_2$} & \begin{tabular}[c]{@{}l@{}}Surface partial pressure of carbon dioxide in sea water\end{tabular} & [$Pa$]         & \\ 
\bottomrule
\end{tabular}%
\end{table}

It is important to notice that the values for all these variables, were collected on each point in the dataset, including also information on latitude, longitude, Spanish autonomous community, and a label for the classes ``\texttt{Posidonia}'' and ``\texttt{Non-Posidonia}''.

\subsubsection{Supplementary Variables}
\label{sec:othervariables}

As many studies in the literature states that bathymetry, water salinity, temperature, and transparency are relevant to predict seagrass locations, we also included these variables in our dataset. 


The Global Ocean Analysis and Forecast~\cite{lellouche2019quality} (GOAF) is based on the NEMO model~\cite{madec_gurvan_2020_3878122} and it has a resolution at $1/12^{\circ}$, providing 10 days of 3D global ocean forecasts updated daily. It includes daily and monthly mean variables on temperature, salinity, currents, sea level, among others\footnote{Available at: \url{https://t.ly/heWA1}}.
Despite the GOAF affords 19 variables, we focused on the two main variables stated by~\cite{bertelli2022use}, which are salinity and temperature, they were collected considering monthly average values for an entire year. However as the GOAF available data ranges from November 2020 to January 2024, we collected data from January 2021 until December 2021.




The European Marine Observation and Data Network (EMODnet) programme is designed to assemble existing, but fragmented and partly inaccessible, marine data and to create contiguous and publicly available information layers which are interoperable and free of restrictions on use, and which encompass whole marine basins~\cite{moses2021mapping}. The EMODnet Digital Bathymetry\footnote{Available at: \url{https://t.ly/mxefH}} is a multilayer bathymetric based upon a collection of surveys, Composite Digital Terrain Model, and Satellite Derived Bathymetry data, with a resolution of $1/16^{\circ}$. As bathymetry is also commonly used as a predictor variable for seagrasses~\cite{bertelli2022use}, we incorporated it into our dataset.


Water transparency is another key indicators for water quality assessment. The data was provided by the Joint Research Centre (JRC) from the satellite sensor MODIS-Aqua Kd490 climatology\footnote{Available at: \url{https://shorturl.at/ahrvV}}. It indicates how deep the sunlight can penetrate in the water, depending on the amount of particles in the water. Particles may be non-living, e.g. sediment from erosion or other dissolved material or living e.g. phytoplankton, microscopic algae. 

The diffuse attenuation coefficient $K_d490$ measures the light penetration in the water column at the blue-green wavelengths (ca. 490 nm). It represents a good indicator of water transparency resulting from the combined action of absorption and backscattering by the water constituents, and the structure of the surrounding light field. These additional variables can be visualized in Table~\ref{tab:dataset_variables2}.

\begin{table}[t]
\caption{Additional physical variables added to the dataset. \ref{tab:dataset_variables}.}
\label{tab:dataset_variables2}
\begin{tabular}{llcc}
\toprule
\textbf{Var}   & \textbf{Description}                                                        & \textbf{Unit} & \textbf{Source} \\ 
\midrule

\texttt{$so$}     & Sea water salinity     & [$psu$] &  \multirow{2}{*}{\begin{tabular}[c]{@{}c@{}} GOAF \end{tabular}}   \\
\texttt{$thetao$} & Sea water potential temperature & [$^\circ C$] & \\ 
\texttt{$depth$}     & \begin{tabular}[c]{@{}l@{}}Measurement of the sea depth \end{tabular}               & [m] &  EMODnet \\
\texttt{$transp$}     & Water transparency
           & [$K_d490$ m$^{-1}$] &  JRC \\ 
\bottomrule
\end{tabular}%
\end{table}

Following the creation of this dataset, our focus transitioned to process the data so, the next section presents the methods and tools used to process and clean the dataset.

\subsection{Data Preprocessing}

Data preprocessing is a fundamental phase in the machine learning pipeline, especially when dealing with classification tasks. This step aims to transform the raw, heterogeneous data into a refined, standardized format that machine learning algorithms can digest. The overall goal is to provide these algorithms with a database that not only speeds up their understanding, but also increases their performance, paving the way for more accurate and reliable predictions. This phase starts with cleaning the raw data by removing missing/null values and outliers. It is also part of this phase to transform the data in a way that the entire dataset can be on the same scale. 

\subsubsection{Data Cleanup}

In real-world datasets, missing values are a common occurrence and can significantly impact model performance. Techniques such as imputation, in which missing values are replaced by statistically derived estimates, or the removal of incomplete instances are employed. Furthermore, identifying and dealing with outliers is crucial to ensuring the robustness of the model. 

As the dataset has a significant number of points, for dealing with the missing values (null values), we decided removing these points, meaning that, for any point in the dataset, if any variable present in Table~\ref{tab:dataset_variables} or Table~\ref{tab:dataset_variables2} does not have a value, that point is discarded. The data points with null values were removed from the dataset by using the Pandas function \texttt{dropna}\footnote{Documentation: \url{https://www.w3schools.com/python/pandas/ref_df_dropna.asp}}.

The outliers were removed by applying the Z-Score method. In this method, each data point is assigned a Z-Score, calculated as the number of standard deviations it is from the dataset mean. The Z-Score provides a standardized measure of how unusual or extreme a data point is within the distribution. We used its implementation in \texttt{SciPy}~\cite{virtanen2020scipy} python library, with the threshold set to 3, indicating that data points beyond three standard deviations from the mean were considered outliers and therefore discarded. Finally, the ``clean'' version of the dataset has 174 columns and 6,020 rows. The total amount of points with and without \textit{P. oceanica} in each Spanish autonomous community is summarized in Table~\ref{tab:points}.

\begin{table}[t]
\centering
\begin{tabular}{lcc}
\hline
\multicolumn{1}{l}{\textbf{Location}} & \multicolumn{1}{c}{\textbf{\textit{P. oceanica}}} & \textbf{Non \textit{P. oceanica}} \\ \hline
Catalonia        & 940          & 961          \\
Valencia         & 996           & 1,108          \\
Murcia           & 471           & 532           \\
Balearic Islands & 303           & 395           \\
Andalucia        & 154           & 160           \\ \hline
\textbf{Total}   & \textbf{2,864} & \textbf{3,156} \\ \hline
\end{tabular}
\caption{Total number of random points generated for each Spanish autonomous community.}
\label{tab:points}
\end{table}

\subsubsection{Data Scaling}

Scaling a dataset is essential when its variables has different units or magnitudes. Scaling or normalization help to avoid distortions that may arise due to numerical disparities between variables. This is particularly relevant for algorithms that rely on distance metrics, such as SVC or KNN, ensuring that each feature contributes proportionately to the model learning process.

So, for scaling the dataset, we used the \texttt{MinMaxScaler} algorithm implemented in \texttt{scikit-learn} library. It specifically transforms data by scaling it to a specified range between 0 and 1, such that all variables values will be in the range [0, 1]. This method operates by subtracting the minimum value in a feature and then dividing by the range (the difference between the maximum and minimum values).

\subsection{Model Development}
\label{sec:model}

Our results were developed on top of \texttt{scikit-learn}~\cite{scikit-learn}, a Python library that provides state-of-the-art implementations of machine learning algorithms. In machine learning, every instance of a particular dataset is represented by a set of variables (or features). If instances are given with known labels (i.e. the corresponding correct outputs) then the learning scheme is known as supervised~\cite{muhammad2015supervised}. 

As stated in Section~\ref{sec:dataset}, our dataset contains a label indicating the presence or absence of \textit{P. oceanica} and the variables referring to biogeochemical information, shown in Table~\ref{tab:dataset_variables}, as well as, the other variables, presented in Table~\ref{tab:dataset_variables2}. So, we implemented and tested a set of supervised machine learning algorithms, reported in the following section.

\subsubsection{Supervised Machine Learning Models}

\textbf{Supervised machine learning} algorithms form a basis of predictive modeling, leveraging labeled datasets to learn and make predictions based on input features. In this type of automatic learning, the algorithms are trained on historical data where input variables are paired with corresponding output labels. The main goal is developing a map function that can accurately predict output labels for new and unseen data. During the development of the model, we tested all \texttt{scikit-learn} supervised algorithms; however, in this paper, we only present the 6 methods with the best prediction accuracy, emphasizing the model with the best results. 


\textbf{Decision Trees (DT)}~\cite{quinlan1986induction} are a supervised learning method used for classification and regression. Its goal is to create a model that predicts the value of a target variable by learning simple decision rules inferred from the data features. The algorithm creates a tree, finding for each node the variable that will yield the largest information gain. Trees are grown to their maximum size and then a pruning step is usually applied to improve the ability of the tree to generalize to unseen data. The generated trees are converted into sets of if-then rules. The accuracy of each rule is then evaluated to determine the order in which they should be applied. 


\textbf{Random Forests}. Significant improvements in classification accuracy have resulted from growing an ensemble of decision trees and letting them vote for the most popular. In order to grow these ensembles, random vectors are generated that govern the growth of each tree in the ensemble. After a large number of trees is generated, they vote for the most popular. Random Forests (RF) are a combination of tree predictors such that each tree depends on the values of a random vector sampled independently and with the same distribution for all trees in the forest~\cite{breiman2001random}.


\textbf{AdaBoost}. The core principle of AdaBoost~\cite{freund1997decision} is to fit a sequence of weak learners (i.e., models that are only slightly better than random guessing, such as small decision trees) on repeatedly modified versions of the data. The predictions from all of them are then combined through a weighted majority vote (or sum) to produce the final prediction. The data modifications at each so-called boosting iteration consists of applying weights $w_1, w_2, ..., w_n$ to each of the training samples. Initially, those weights are all set to $w_i = 1/N$, so that the first step simply trains a weak learner on the original data. For each successive iteration, the sample weights are individually modified and the learning algorithm is reapplied to the reweighted data. At a given step, those training examples that were incorrectly predicted by the boosted model induced at the previous step have their weights increased, whereas the weights are decreased for those that were predicted correctly. As iterations proceed, examples that are difficult to predict receive ever-increasing influence. Each subsequent weak learner is thereby forced to concentrate on the examples that are missed by the previous ones in the sequence~\cite{hastie2009elements}.


\textbf{Neighbors-based classification}~\cite{cover1967nearest} is a type of instance-based learning that stores instances of the training data. Classification is computed from a simple majority vote of the nearest neighbors of each point: a query point is assigned the data class which has the most representatives within the \texttt{K} nearest neighbors (KNN) of the point. During the training phase, the KNN algorithm stores the entire training dataset as a reference. When making predictions, it calculates the distance between the input data point and all the training examples, using a chosen distance metric such as Euclidean distance. Next, the algorithm identifies the \texttt{K} nearest neighbors to the input data point based on their distances, assigning the most common class label among these neighbors. 


\textbf{A support vector machine} constructs a hyper-plane or set of hyper-planes in a high or infinite dimensional space, which can be used for classification, regression or other tasks. Intuitively, a good separation is achieved by the hyper-plane that has the largest distance to the nearest training data points of any class (so-called functional margin), since in general the larger the margin the lower the generalization error of the classifier~\cite{boser1992training}. Support Vector Classifiers (SVC) can be effective in high dimensional spaces, even when the number of dimensions is greater than the number of samples. Also, it is memory efficient, as it uses a subset of training points in the decision function (support vectors).


\textbf{Multi-layer Perceptron (MLP)}~\cite{rumelhart1986learning} is a back-propagation neural network algorithm. Characterized by its layered structure, the MLP consists of an input layer, one or more hidden layers, and an output layer. Each layer comprises interconnected nodes, or neurons, with each connection possessing an associated weight. The learning process occurs through a series of forward and backward passes, where input data is propagated through the network, and the model adjusts its weights based on the error between predicted and actual outcomes. The algorithm repeatedly adjusts the weights of the connections in the network so as to minimize a measure of the difference between the actual output vector of the net and the desired output vector. As a result of the weight adjustments, internal `hidden' neurons, which are not part of the input or output, come to represent important features of the task domain, and the regularities in the task are captured by the interactions of these neurons.

\subsubsection{Model Evaluation}
\label{sec:model_evaluation}

Evaluating the performance of a machine learning classification algorithm is a fundamental phase in the model development. This step allows to understand the effectiveness of the algorithm and its potential for generalization beyond the limits of the training data. Effective evaluation not only assesses the extent to which the model has learned patterns from a given dataset, but also sheds light on its adaptability and predictive capabilities when faced with new and unseen instances.


For evaluating the models, we first split the dataset into training and test sets, a pivotal step for assessing the model's ability to generalize to unseen data. It also ensures that the performance of the model is not overly influenced by the data on which it was trained and can effectively adapt to new, unseen points. At this point, we used the \texttt{train\_test\_split} function of \texttt{scikit-learn}, which takes the dataset as input considering its variables as $X$ and the label as $y$. The function returns four subsets: the training data (\texttt{X\_train}, \texttt{y\_train}) and the testing data (\texttt{X\_test}, \texttt{y\_test}). The split was defined to 0.5, so it returned half of the dataset for training the classifiers and half as the test dataset. 

With the training dataset, we applied the \texttt{scikit-learn} \texttt{KFold} function which operates by partitioning the dataset into K equally sized folds, ensuring that each observation appears in exactly one training and one validation set. The function returns an iterable of indices, representing the split positions for each fold. This allows to iterate through the folds, training and evaluating the model K times, each time using a different fold as the validation set. For all prediction tests we defined a 5-fold cross-validation, meaning that the dataset was divided into five subsets. The model was trained on four folds and validated on the remaining one, and this process was repeated five times, with each fold acting as the validation set exactly once. This procedure was applied for all classifiers reported in Section~\ref{sec:results}.


The evaluation of machine learning performance is performed through metrics. These metrics, rooted in statistical and mathematical principles, provide a comprehensive assessment of a model's predictive accuracy, precision, recall, and other essential facets. 
These metrics form the foundation for making informed decisions about model deployment, optimization, and the overall success of machine learning endeavors. We evaluated the models considering the following metrics.

\textbf{Confusion Matrix} is a tool for evaluating classification models. It provides a detailed breakdown of the model's predictions, including true positives, true negatives, false positives, and false negatives. \textbf{Accuracy} is a straightforward measure that calculates the ratio of correctly predicted instances to the total number of instances. While easy to interpret, accuracy alone may be insufficient, especially in the presence of imbalanced datasets. To compute the Confusion Matrices in this paper, we called the \texttt{scikit-learn} function \texttt{confusion\_matrix}.

\textbf{Precision}, \textbf{Recall}, and \textbf{F1-score} offer a more nuanced evaluation, especially in imbalanced scenarios. Precision measures the proportion of correctly predicted positive instances among all instances predicted as positive, while recall gauges the proportion of correctly predicted positive instances among all actual positive instances. The F1-score is the harmonic mean of precision and recall, providing a balanced assessment. These values are returned by the \texttt{scikit-learn} function \texttt{classification\_report}.

\textbf{Support} refers to the number of instances or samples in each class. It provides insights into the distribution of data across different classes. In classification reports or confusion matrices, you can observe the support for each class, helping to identify whether the model is trained on a balanced or imbalanced dataset. It is particularly crucial when dealing with imbalanced classes, as it aids in assessing the significance and representation of each class in the dataset.

\textbf{Macro Average} calculates the average performance metric (such as precision, recall, or F1-score) across all classes without considering class imbalance. Each class is given equal weight in the calculation, irrespective of the number of instances in each class. This approach provides a uniform evaluation across different classes, making it particularly useful when all classes are considered equally important. \textbf{Weighted Average}, on the other hand, considers the class imbalance by calculating the average metric while taking into account the number of instances in each class. Larger classes have a proportionally greater impact on the average, acknowledging their significance in the overall performance of the model. Weighted average is beneficial in scenarios where class imbalance is prevalent, as it ensures that the evaluation metric is more reflective of the model's performance on the majority classes.

\textbf{ROC Curve} (Receiver Operating Characteristic) serves as a visual representation in binary classification, representing the trade-off between sensitivity and specificity at various classification thresholds. Sensitivity, which indicates the correct identification of positive instances, is plotted against the false positive rate, representing the misclassification of negative instances. This curve provides insights into a model's performance at different decision thresholds, with the Area Under the ROC Curve (AUC-ROC) offering a scalar measure of overall discriminative ability. Valuable in imbalanced data sets or when considering variable costs of misclassification, the ROC curve allows practitioners to evaluate and compare the performance of classification models, guiding decisions about model suitability for specific tasks.

\textbf{Features Importance} provides insights into the importance of each variable in making predictions. It helps users understand which features contribute the most to the predictive power of the model. For example, in a Random Forest, each decision tree in the is trained on a random subset of the data, and at each split in a tree, a subset of features is considered. The feature importance calculates the relative importance of each feature based on how frequently a feature is used to split the data across all the trees in the forest. Features with higher importance values are considered more influential in the model's decision-making process. A higher value indicates that the feature is more effective at reducing impurity or increasing information gain when used in tree splits. The feature importance values represent the relative importance of features in a machine learning model and are unitless. The values are normalized, meaning they sum to 1, allowing for a comparison of the relative contributions of different features to the model's predictions.


\section{Conclusions}
\label{sec:conclusion}

This study provides compelling evidence of the significant impact of \textit{P. oceanica} on water conditions in coastal ecosystems at a global scale. Using augmented research methodology we curated an expansive dataset comprising 174 biogeochemical and physical properties, aggregating data from several open sources of all autonomous communities in Spain bathed by the Mediterranean Sea. With this dataset, we achieved a precision rate of up to 90\% in determining the plant exact location. This method can be utilized to identify seagrass from indirect sources in other locations.

Our machine learning analysis revealed a robust correlation between the location of \textit{P. oceanica} meadows and the water biogeochemical properties, allowing to identify key indicators that correlate with the presence of the seagrass. Notably, carbon-related variables such as net biomass production, downward surface mass flux and surface partial pressure of carbon dioxide, exhibited distinct alterations in areas with \textit{P. oceanica}, suggesting their potential utility as indirect indicators of seagrass meadow locations and confirming that this seagrass is responsible for carbon sequestration. Furthermore, it was possible to perceive an increase on oxygen production in the areas with \textit{P. oceanica}, because during the process of photosynthesis, seagrass absorbs carbon dioxide from the water column and convert it into oxygen. 

The findings of our study offer valuable insights that can inform the development of targeted conservation strategies for preserving marine ecosystems. By considering several variables associated with the presence of seagrass, such strategies can contribute to the long-term sustainability of coastal habitats and the preservation of marine biodiversity. This is particularly important considering the growing recognition of the crucial role that coastal ecosystems play in regulating the climate. \textit{P. oceanica}, in particular, stands out for its ability to sequester carbon dioxide and mitigate the impacts of climate change, highlighting its importance as a natural climate solution.


\subsection*{Data and Code Availability}

The dataset and the python code for processing it, training and evaluating the models, are available at \url{https://github.com/celiotrois/posidonia_biogeochemical}.

\bmhead{Acknowledgements}

The authors are grateful to Dr. Elena Díaz Almela for inspiring discussions and support.
Authors acknowledge project TED2021-131420B-I00 funded by MCIN/AEI/10.13039/501100011033 and by European Union NextGenerationEU/PRTR.

\bibliography{sn-bibliography}

\end{document}